# Sub-natural width of transparency window in $^{85}$Rb vapor with D$_2$ transition

S. M. IFTIQUAR

Department of Physics, Indian Institute of Science, Bangalore-560012, India.

Email: smiftiquar@gmail.com

Electromagnetically induced transparency (EIT) is a quantum interference phenomenon [1] and since its discovery a lot of interest has been drawn on its potential application in slowing down speed of light with EIT. Because of its sub-natural line width and quantum mechanical nature it is potential candidate for quantum information processing [2]. It has been shown that anomalous dispersion profile near resonance condition increases with reduction in width of transparency window, which enhances capability of optical group velocity reduction, however reduced transmission width reduced spectral bandwidth. It has recently been demonstrated that light can be slowed and stopped with the help of EIT and optical information processing can be carried out efficiently [2]. In most of such experiments $^{87}$Rb has been employed because of its favorable energy level configuration. However here we explore characteristics of $^{85}$Rb atoms and its related EIT spectra. EIT can be observed in many different energy level configuration, like $\Lambda$, V or ladder configuration etc.

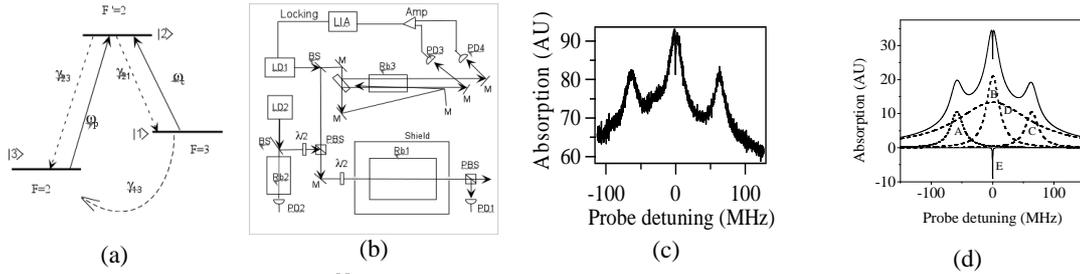

Fig 1. a) Energy level diagram of $^{85}$Rb ground states F=3, 2 and excited state F'=2 denoted as |1>, |3> and |2>. Pair of solid and dotted arrows indicate laser induced excitation and spontaneous decay channels. (b) Experimental setup for EIT. (c) Typical $^{85}$Rb EIT spectra, as observed in experiment. (d) Theoretical result of $^{85}$Rb EIT spectra and deconvolution of its Lorentzian components.

Density matrix equation of motion for $\Lambda$ system, of figure 1a, can be written as

$$\dot{\rho}_{21} = -(\omega_{21} + \gamma_{21})\rho_{21} + i\Omega_1(\rho_{22} - \rho_{11})$$
$$\dot{\rho}_{23} = -(\omega_{23} + (\gamma_{23} + f))\rho_{23} + i\Omega_3(\rho_{22} - \rho_{33})$$
(1)

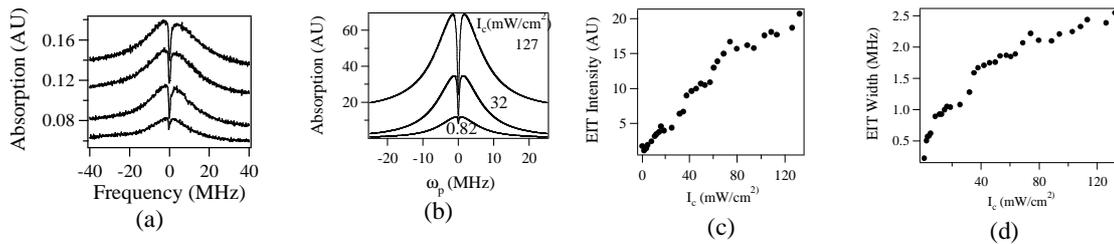

Fig 2: $^{85}$Rb EIT spectra (a) Experimentally observed, (b) theoretical simulation of EIT spectra for different control laser intensity. (c) Variation of integrated intensity of EIT signal and (d) variation of Lorentzian half-width of EIT spectra with increasing control laser intensity.

f indicates line-width broadening effect like power broadening, collisional broadening, transit time broadening etc $\gamma_{21}$ atomic decay rate from $|2\rangle \to |1\rangle$ without any environmental influence. In rotating wave approximation this leads to (in electric dipole approximation) EIT, which, in small frequency range (around Zeeman quantum number $m_F=m_{F'}=0$) can be approximated as

$$T(\omega)_{23} = K \frac{(\gamma_{23}^0 + f)}{(\omega_{23} - \omega_p)^2 + (\gamma_{23}^0 + f)^2} \qquad (2)$$

where $\omega_{23}$ is atomic resonance frequency for transition $|2\rangle \to |3\rangle$, $\gamma_{23}^0$ decay constant corresponding to Zeeman quantum number $m_F=m_{F'}=0$, K is a constant.

Experimental setup is shown in figure 1b. Both the lasers are in continuous wave (cw) operation. Control laser is locked to saturation peak F=3 $\to$ F'=2 and probe laser is frequency scanned from F=2 to all excited state of $D_2$ transition at 5 Hz scan rate. Both the lasers are orthogonally linearly polarized. The observed EIT spectra is shown in figure 1c. Figure 1d is theoretical trace that shows the components of spectra contributing to the EIT signal 1c. Figure 2a shows experimentally observed EIT spectra at various control laser intensity; increases for upper traces. Figure 2b is theoretical EIT spectra. A deconvolution of the EIT spectra shows that the whole spectra is formed by convolution of several Lorentzian spectra (figure 1d), A, B, C, D, E. The frequency separation between A and B is $2 \times \omega_{1'2'}$, and $\omega_{BC} = \omega_{2'3'}$. EIT occurs (transmission E) at B, whose frequency is $\omega_{22'}$. The broad background absorption D is Doppler broadened absorption while peaks B, C are absorption due to excited state hyperfine levels, F'=2,3 respectively.

In figure 2a, 2b, shows Autler-Townes doublets becomes more widely separated at higher laser intensity. Figure 2c shows that intensity of EIT spectra increases with control laser intensity, with a simultaneous increase in Lorentzian width of the EIT spectra, figure 2d. We have obtained similar EIT spectra with $\sigma^+ - \sigma^-$ laser beams as well.

In conclusion, $^{85}$Rb EIT spectra exhibits good EIT spectra, and the EIT does not disappear at optical intensity higher than saturation intensity, as suggested in literature [3]. The result shows the strength of EIT signal increases at higher optical power, although the transmission width reduces at the same time.

In acknowledgements, this work was supported by the Council of Scientific and Industrial Research of India.